\documentclass{procstyle}

\usepackage{refmerge} 

\input definitions
\input myBabarsym

\def\ggg{g}

\begin{document}

\title{The Hadronic Contribution to \boldmath$(\ggg-2)_\mu$}

\author{Andreas H\"ocker}

\address{Laboratoire de l'Acc\'el\'erateur Lin\'eaire,\\
         IN2P3-CNRS et Universit\'e Paris-Sud --
         B\^at. 200, BP34 -- F-91898 Orsay, France\\
         E-mail: hoecker@lal.in2p3.fr}

\twocolumn[\maketitle\abstract{
The evaluation of the hadronic contribution to the muon magnetic 
anomaly $a_\mu$ is reviewed, including a new estimate using 
precise results on the $\pip\pim$ spectral function from the 
KLOE Collaboration. It is found that the KLOE data confirm to
some extent the previous $\epem$ annihilation data in this channel,
and accentuate the disagreement with the isospin-breaking-corrected
spectral function from $\taum\to\pim\piz\nut$ decays. Correcting
for the empirical difference in the mass of the charged and
the neutral $\rho$ locally improves, but does not resolve this 
discrepancy. A preliminary reevaluation (including the KLOE data) 
of the \epem-based Standard Model prediction of $a_\mu$ results in 
a deviation of 2.7 standard deviations from the BNL measurement.}]

%
%

\section{Introduction}

Hadronic vacuum polarization (HVP) in the photon propagator plays an 
important role in many precision tests of the Standard Model. 
%
%
This is the case for the muon anomalous magnetic moment 
$a_\mu\equiv(g_\mu -2)/2$, where the HVP component is the leading 
contributor to the uncertainty of the Standard Model prediction.
The HVP contribution is computed by means of a dispersion relation
as an integral over experimentally determined spectral functions.
It is the property of this dispersion relation that the $\pi\pi$ 
spectral function provides the major part of the total HVP
contribution, so that the experimental effort focuses on this channel.

Spectral functions are directly obtained from the  cross sections of \epem
annihilation into hadrons. The accuracy of the calculations has therefore
followed the progress in the quality of the corresponding data\cite{eidelman}.
Because the latter were not always suitable, it was deemed necessary to 
resort to other sources of information. One such possibility was the 
use of the vector spectral functions\cite{adh} derived from the study 
of hadronic $\tau$ decays\cite{aleph_vsf} for the energy range less 
than $m_\tau\simeq1.8\gevcc$. For this purpose, 
the isospin rotation that leads from 
the charged $\tau$ to the neutral \epem  final state has to be 
thoroughly corrected for isospin-breaking effects.

Also, it was demonstrated that essentially perturbative QCD could be 
applied to energy scales as low as $1$--$2\gev$\cite{aleph_asf,opal_alphas},
thus offering a way to replace poor \epem data in some energy regions 
by a reliable and precise theoretical 
prescription\cite{martin,dh97,steinhauser,erler,groote,dh98}. 

Detailed reanalyses including all available experimental data have been
published in Refs.\cite{dehz03,dehz,teubner} (see also the preliminary 
results given in Refs.\cite{yndurain,jegerlehner}), taking advantage 
of precise results in the $\pi\pi$ channel from the CMD-2 
experiment\cite{cmd2_new} and from the ALEPH analysis of $\tau$ 
decays\cite{aleph_new}, and benefiting from a more complete
treatment of isospin-breaking corrections\cite{ecker1,ecker2}. 
It was found that the \epem  and the isospin-breaking-corrected 
$\tau$ spectral functions do not agree within their respective 
uncertainties, thus leading to inconsistent predictions for the 
lowest-order hadronic contribution to $a_\mu$. The 
dominant contribution to the discrepancy stems from the $\pi\pi$ channel
with a difference of $(-11.9\pm6.4_{\rm exp}\pm2.4_{\rm rad}\pm2.6_{\rm SU(2)}
\,(\pm7.3_{\rm total}))\tmten$, and a more significant energy-dependent
deviation.\footnote
{
	The problem between $\tau$ and \epem data is more noticable
	when comparing the $\tau\to\pi\piz\nu$ branching fraction
	with the prediction obtained from integrating the 
	corresponding isospin-breaking-corrected \epem
	spectral function. Here, the function under the integrand
	is less selective than it is the case for the HVP contribution
  	to $a_\mu$, leading to a 
	discrepancy of 2.9 standard deviations\cite{dehz03}.
}
When compared to the world average of the muon magnetic anomaly
measurements,
dominated by the results from the BNL experiment\cite{bnl_2004},
\beq
\label{eq:bnlexp}
	a_\mu \:=\: (11\,659\,208.0 \pm 5.8)\tmten~,
\eeq
the respective \epem  and $\tau$-based predictions disagreed at 
the level of 2.5 and 1.3 standard deviations, when adding 
experimental and theoretical errors in quadrature.

This summer, new data on the $\pi\pi$ spectral function
in the mass region between $0.60$ and $0.97\gevcc$ were presented by
the KLOE Collaboration\cite{kloe_pipi}, using the---for the purpose
of precision measurements---innovative technique 
of the radiative return\cite{isr}. The statistical precision 
of these data by far outperforms the Novosibirsk sample, but the
systematic errors are about twice as large as those obtained by 
CMD-2. New data using the same technique have been published by
the \babar\  Collaboration\cite{babar_3pi} on the 
$\pip\pim\piz$ final state. They unveil a larger cross sections and 
a resonant peak at around $1.6\gevcc$ that was missed by the 
previous DM2 measurement\cite{dm2_3pi}. The \babar\   data are not 
(yet) used in the preliminary reevaluation of the lowest-order HVP 
contribution given here.\footnote
{
  	The correction to \amuhadLO\  when using the \babar\ 
	data for this mode is of the order of $+1\tmten$.
}

%
%

\section{Muon Magnetic Anomaly}
\label{sec:anomaly}

It is convenient to separate the Standard Model (SM) prediction for the
anomalous magnetic moment of the muon
into its different contributions,
\beq
    a_\mu^{\rm SM} \:=\: a_\mu^{\rm QED} + a_\mu^{\rm had} +
                             a_\mu^{\rm weak}~,
\eeq
with
\beq
 a_\mu^{\rm had} \:=\: a_\mu^{\rm had,LO} + a_\mu^{\rm had,HO}
           + a_\mu^{\rm had,LBL}~,
\eeq
and where $a_\mu^{\rm QED}=(11\,658\,472.0\pm0.2)\tmten$ is 
the pure electromagnetic contribution (see\cite{hughes,cm} and references 
therein), \amuhadLO\ is the lowest-order HVP contribution, 
$a_\mu^{\rm had,HO}=(-10.0\pm0.6)\tmten$ 
is the corresponding higher-order part\cite{krause2,adh}, 
and $a_\mu^{\rm weak}=(15.4\pm0.1\pm0.2)\tmten$,
where the first error is the hadronic uncertainty and the second
is due to the Higgs mass range, accounts for corrections due to
exchange of the weakly interacting bosons up to two loops\cite{amuweak}. 
For the light-by-light (LBL) scattering part, $a_\mu^{\rm had,LBL}$,
we use the value $(12.0\pm3.5)\tmten$ from the latest 
evaluation\cite{lbl_mv}, slightly corrected for the missing 
contribution from (mainly) the pion box.

Owing to unitarity and to the analyticity of the vacuum-polarization 
function, the lowest order HVP contribution to $a_\mu$ can be computed 
\via\ the dispersion integral\cite{rafael}
\beq
\label{eq_int_amu}
    a_\mu^{\rm had,LO} \:=\: 
           \frac{\alpha^2(0)}{3\pi^2}
           \intl_{4m_\pi^2}^\infty\!\!ds\,\frac{K(s)}{s}R^{(0)}(s)~,
\eeq
where $K(s)$ is a well-known QED kernel, and
$R^{(0)}(s)$ denotes the ratio of the ``bare'' cross
section for \epem annihilation into hadrons to the pointlike muon-pair cross
section. The bare cross section is defined as the measured cross section
corrected for initial-state radiation, electron-vertex loop contributions
and vacuum-polarization effects in the photon propagator. However, photon 
radiation in the final state is included in the bare cross section 
defined here. The reason for using the bare (\ie, lowest order) 
cross section is that a full treatment of higher orders is anyhow 
needed at the level of $a_\mu$, so that the use of the ``dressed''
cross section would entail the risk of double-counting some of the 
higher-order contributions.

The function $K(s)\sim1/s$ in Eq.~(\ref{eq_int_amu}) gives a strong 
weight to the low-energy part of the integral. About 91\pc\ of the 
total contribution to \amuhadLO\  is accumulated at center-of-mass
energies $\sqrt{s}$ below $1.8\gev$ and 73\pc\ of \amuhadLO\ is covered 
by the $\pi\pi$ final state, which is dominated by the $\rho(770)$ 
resonance. 

%
%
\section{The Input Data}

\begin{figure*}
  \centerline{
	  \epsfxsize7cm\epsffile{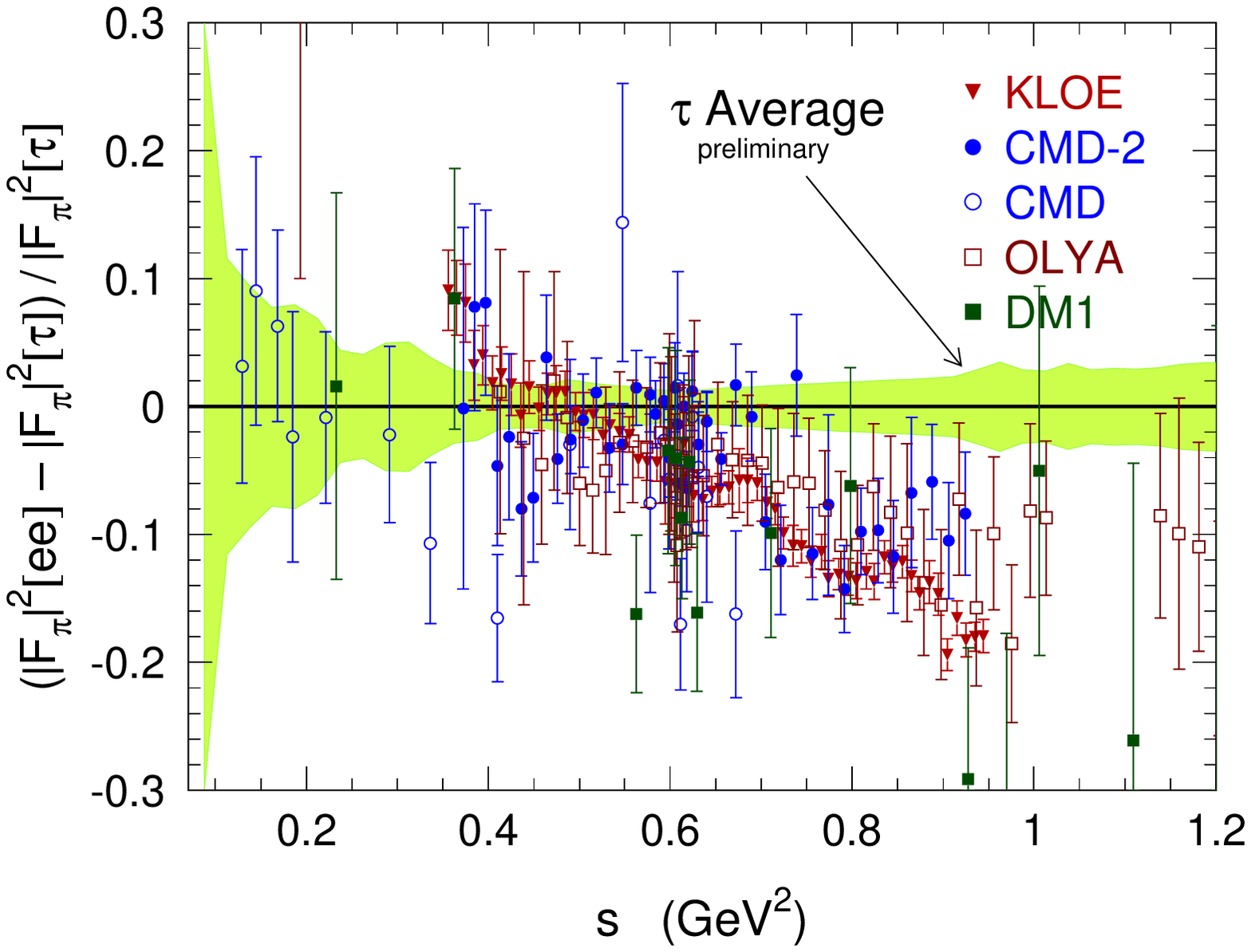}
          \hspace{0.05cm}
	  \epsfxsize7cm\epsffile{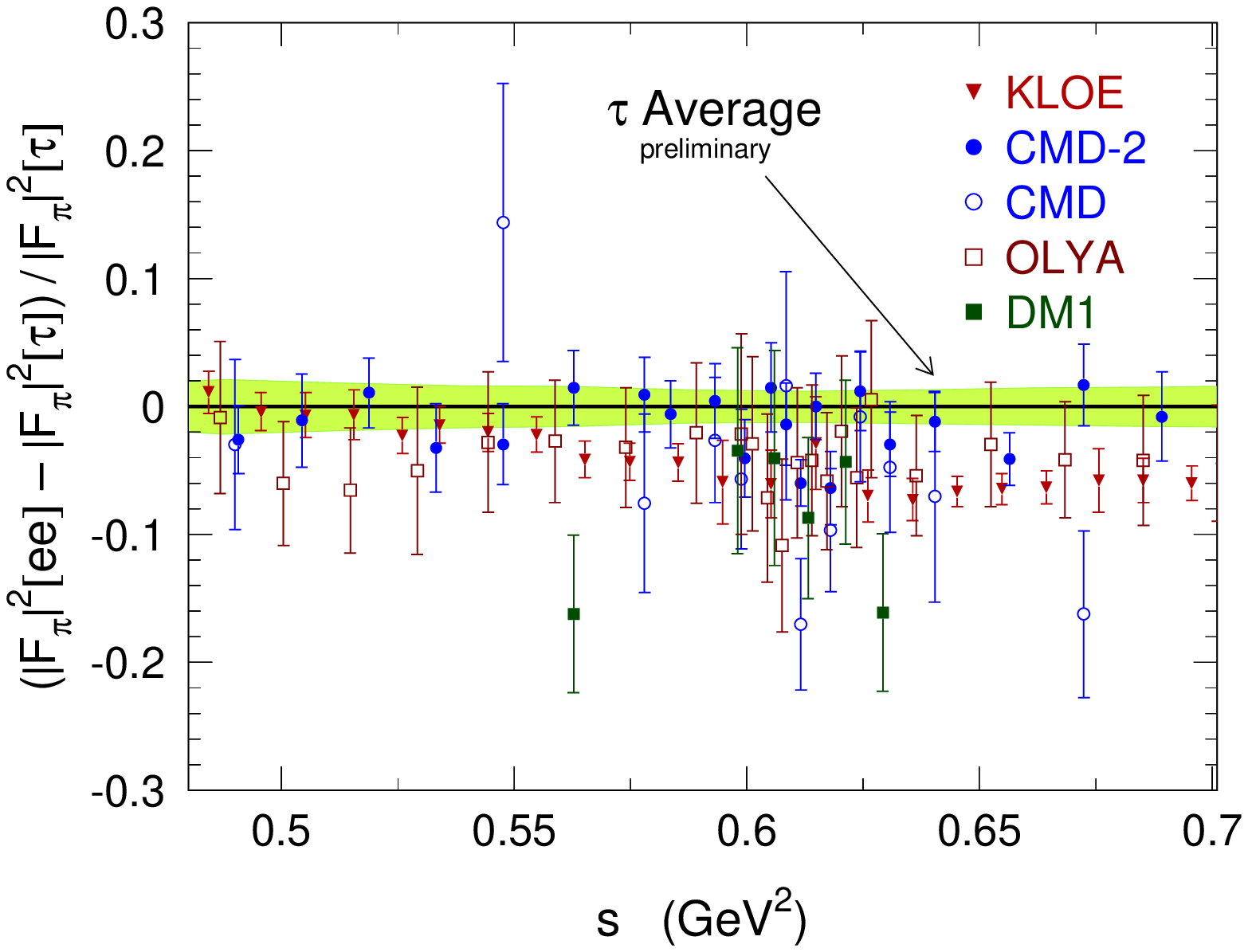}
	     }
  \caption[.]{\label{fig:2pi_comp}
	Relative comparison of the $\pi^+\pi^-$ spectral functions
    	from \epem-annihilation data and isospin-breaking-corrected 
	$\tau$ data, expressed as a ratio to the $\tau$ spectral function.
	The shaded band indicates the errors of the $\tau$ data.
        The \epem  data are from KLOE\cite{kloe_pipi}, 
	CMD-2\cite{cmd2_new}, CMD, OLYA and DM1 
        (references given in Ref.\cite{dehz03}).
        The right hand plot emphasizes the region of the $\rho$ 
        peak.}  
\end{figure*}

A detailed compilation of all the experimental data used in the 
evaluation of the dispersion integral~(\ref{eq_int_amu}) is provided in 
Refs.\cite{dehz,dehz03}. Also discussed therein is the corrective 
treatment of radiative effects applied to some of the measurements.
The $\tau$ spectral function is obtained by averaging the results from
ALEPH\cite{aleph_vsf}, CLEO\cite{cleo_2pi} and OPAL\cite{opal_2pi},
which exhibit satisfactory mutual agreement.

A comparison of the $e^+e^-\rightarrow\pip\pim$ data and the 
corresponding $\tau$ spectral function, represented as a point-by-point ratio 
to the $\tau$ spectral function  is given in Fig.~\ref{fig:2pi_comp}. 
Several observations can be made. 
\bei

\item 	A significant discrepancy, mainly above the $\rho$ peak
	is found between $\tau$ and the \epem data from CMD-2 as well	
	as older data from OLYA.

\item	Overall, the KLOE data confirm the trend exhibited by the 
	other \epem data.

\item	Some disagreement between KLOE and CMD-2
	occurs on the low mass side (KLOE data are large),
  	on the $\rho$ peak (KLOE below CMD-2) as well as
	on the high mass side (KLOE data are low).

\eei

At this stage, the $\tau$ spectral function  has not been corrected 
for a possible $\rho^- - \rho^0$ mass and width 
splitting\cite{gj,md_tausf}.
In contrast to earlier experimental\cite{aleph_vsf} and 
theoretical results\cite{bijnens}, a combined pion form 
factor fit\cite{md_tausf} to the new precise data on \epem 
and $\tau$ spectral functions leads to 
$m_{\rho^-}-m_{\rho^0}=(2.3\pm0.8)\mevcc$, while no 
significant width splitting$^{\,}$\footnote
{
  Note that if the mass difference is to be taken as an 
  experimental fact, a larger width difference would be
  expected. Using a chiral model of the $\rho$ 
  resonance\cite{pich-portoles,ecker1,ecker2}, one has
  $
  \Gamma_{\rho^0}
  =\Gamma_{\rho^-}(m_{\rho^0}/m_{\rho^-})^3 
                  (\beta_0/\beta_-)^3
   +\Delta \Gamma_{\rm EM}	
  $,
  where $\Delta \Gamma_{\rm EM}$ is the width difference from 
  electromagnetic decays. This leads to a total width difference 
  of $(2.1 \pm 0.5)\mevcc$ that is marginally consistent with the 
  observed value\cite{md_tausf}.
} is 
observed within the fit error of $1.7\mevcc$.

Considering the mass splitting in the isospin-breaking correction 
of the $\tau$ spectral function  tends to locally improve (though 
not restore) the agreement between $\tau$ and CMD-2 data, leaving 
an overall normalization discrepancy. Increasing the
$\Gamma_{\rho^-}-\Gamma_{\rho^0}$ width splitting by $+3\mevcc$
improves the agreement between $\tau$ and KLOE data in the 
peak region, while it cannot correct the discrepancies in
the tails. Note that a correction of the mass splitting 
alone would {\em increase} the discrepancy between the $\tau$ 
and \epem-based  results for \amuhadLO.

During the previous evaluations of \amuhadLO, the results using 
respectively the $\tau$ and \epem  data were quoted individually, 
but on the same footing since the \epem-based evaluation was 
dominated by the data from a single experiment (CMD-2).
The confirmation of this discrepancy by KLOE discredits the 
$\tau$-based result for the use in the dispersion integral
until a better understanding of the dynamical origin of the 
observed effect is achieved. This is a challenging problem,
which may itself turn out to be of fundamental importance.

%
%
\section{Results}
\label{sec_results_amu}

\begin{figure}[t]
\centerline{\psfig{file=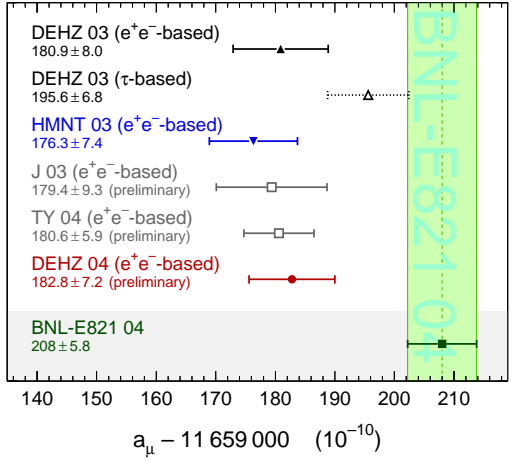,width=70mm}}
\caption[.]{\label{fig:results}
	Comparison of the result~(\ref{eq:smres}) with the 
	BNL measurement\cite{bnl_2004}. Also given
	are our previous estimates\cite{dehz03}, where the 
	triangle with the dotted error bar indicates the 
	$\tau$-based result, as well as the estimates from 
	Refs.\cite{teubner,yndurain,jegerlehner}, not yet including 
	the KLOE data.
	}
\end{figure}
The inclusion of the KLOE $\pi\pi$ data decreases the contribution
from this mode from\cite{dehz03} $(450.2\pm4.9\pm1.6_{\rm rad})\tmten$
to $(448.3\pm4.1\pm1.6_{\rm rad})\tmten$ for the energy interval 
between 0.5 and $1.8\gev$. Note that the additional systematic error
due to radiative effects originates from the energy regions
not covered by the recent KLOE and CMD-2 measurements, where
a full treatment of radiative corrections is applied. 
The preliminary estimate of the integral~(\ref{eq_int_amu}) 
given below includes one 
additional improvement with respect to Ref.\cite{dehz03}:
perturbative QCD is used instead of experimental data in the region 
between $1.8$ and $3.7\gev$, where non-perturbative contributions to 
integrals over
differently weighed spectral functions were found to be small\cite{dh97}.
This results in a reduction of \amuhadLO \  by $-1\tmten$.
All other contributions to the dispersion integral are equal to 
those defined in Ref.\cite{dehz03}.

The \epem-based result for the lowest order hadronic contribution is
\beq
  \amuhadLO =
       (693.4\pm5.3\pm3.5_{\rm rad})\tmten ~,
\eeq
where the second error is due to our treatment of (potentially) 
missing radiative corrections in the older data\cite{dehz}. 
Adding to this the QED, higher-order hadronic, light-by-light 
scattering, and weak contributions given in Section~\ref{sec:anomaly},
one finds
\beqn
\label{eq:smres}
  a_{\mu}^{\rm SM} &=& (11\,659\,182.8
                     \pm 6.3_{\rm had,LO+HO}~~~~~~~~~ \nonumber\\
		   && \hspace{-0.8cm}
                     \pm\, 3.5_{\rm had,LBL} 
                     \pm 0.3_{\rm QED+EW})\tmten~.
\eeqn

This value can be compared to the present measurement~(\ref{eq:bnlexp});
adding all errors in quadrature, the difference between experiment
and theory is
\beq
\label{eq:diffbnltheo}
  a_\mu^{\rm exp}-a_{\mu}^{\rm SM} =
	(25.2\pm9.2)\tmten~,
\eeq
which corresponds to 2.7 ``standard deviations'' (to be interpreted
with care due to the dominance of systematic errors in the SM
prediction).
A graphical comparison of the result~(\ref{eq:smres}) with 
previous evaluations and the experimental value is given in 
Fig.~\ref{fig:results}. 

\section{Conclusion and Perspective}

In spite of the new and precise data on the two-pion
spectral function from the KLOE Collaboration, the lowest
order hadronic vacuum-polarization contribution remains the 
most critical component in the Standard Model prediction
of $a_\mu$. The central piece of information provided by 
the present KLOE data (a reanalysis with 
better quality data and a refined study of systematic
effects can be expected\cite{graziano}) is that 
they confirm the discrepancy between the $\tau$ data and 
\epem  annihilation observed in this channel\cite{dehz03}.
This said, we point out that there also occurs
disagreement between KLOE and CMD-2 data in some 
of the energy regions.

An empirical isospin-breaking correction of the $\rho$ resonance 
lineshape (mass and width) improves but does not
restore the agreement between the two data sets.
It is a consequence of this confirmation that, until the 
CVC puzzle is solved, only \epem  data should 
be used for the evaluation of the dispersion integral.
Doing so, and including the KLOE data, we find that the 
Standard Model prediction of $a_\mu$ differs from the 
experimental value by 2.7 standard deviations.

We are looking forward to the forthcoming results on the 
low- and high-energy two-pion spectral function from the 
CMD-2 Collaboration. These data will help to significantly
reduce the systematic uncertainty due to the
corrective treatment of radiative effects, often omitted 
by part by the previous experiments.

The initial-state-radiation program
of the \babar\  collaboration has already proved its
performance by publishing the spectral function
for $\pip\pim\piz$ (and soon for $2\pip 2\pim$),
while results for the two-pion final state are 
expected.

%
%

\section*{Acknowledgements}

I am indebted to Arkady Vainshtein and Graziano Venanzoni 
for informative discussions, and to Bill Marciano for pointing 
out a mistake in the citation of the QED contribution.
I thank Michel Davier, Simon Eidelman
and Zhiqing Zhang for the fruitful collaboration.


\begin{thebibliography}{99}

\bibitem{eidelman}    	S.~Eidelman and F.~Jegerlehner,
                      	Z. Phys. {\bf C67}, 585 (1995).

\bibitem{adh}         	R.~Alemany, M.~Davier and A.~H\"ocker,
                      	Eur. Phys. J. {\bf C2}, 123 (1998).

\bibitem{aleph_vsf}   	ALEPH Collaboration (R.~Barate \ea),
                      	Z. Phys. {\bf C76}, 15 (1997).

\bibitem{aleph_asf}   	ALEPH Collaboration (R.~Barate \ea),
                      	Eur. J. Phys. {\bf C4}, 409 (1998).

\bibitem{opal_alphas} 	OPAL Collaboration (G. Abbiendi \ea),
                      	Eur. Phys. J. {\bf C7}, 571 (1999).

\bibitem{martin}      	A.D.~Martin and D.~Zeppenfeld,
                      	Phys. Lett. {\bf B345}, 558 (1995).

\bibitem{dh97}        	M.~Davier and A.~H\"ocker,
                      	Phys. Lett. {\bf B419}, 419 (1998).

\bibitem{steinhauser} 	J.H.~K\"uhn and M.~Steinhauser,
                      	Phys. Lett. {\bf B437}, 425 (1998).

\bibitem{erler}         J.~Erler, 
                        Phys. Rev. {\bf D59}, 054008 (1999).

\bibitem{groote}      	S.~Groote \ea,
                      	Phys. Lett. {\bf B440}, 375 (1998).

\bibitem{dh98}        	M.~Davier and A.~H\"ocker,
                      	Phys. Lett. {\bf B435}, 427 (1998).

\bibitem{dehz03}      	M.~Davier, S.~Eidelman, A.~H\"ocker and Z.~Zhang,
                      	Eur. Phys. J. {\bf C31}, 503 (2003).

\bibitem{dehz}        	M.~Davier, S.~Eidelman, A.~H\"ocker and Z.~Zhang,
                      	Eur. Phys. J. {\bf C27}, 497 (2003).

\bibitem{teubner}     	K.~Hagiwara, A.D.~Martin, D.~Nomura and
		      	T.~Teubner, Phys. Lett. {\bf B557}, 69 (2003).

\bibitem{yndurain}    	J.F.~de Troconiz and F.J. Yndurain,
                      	FTUAM-04-02, hep-ph/0402285 (2004).

\bibitem{jegerlehner} 	F.~Jegerlehner, 
                      	DESY-03-189, hep-ph/0312372 (2003).

\bibitem{cmd2_new}    	CMD-2 Collaboration (R.R.~Akhmetshin \ea),
                      	Phys. Lett. {\bf B578}, 285 (2004).

\bibitem{aleph_new}   	M.~Davier and C.~Yuan,
                      	Nucl. Phys. (Proc. Suppl.) {\bf B123}, 47 (2003).

\bibitem{ecker1}      	V.~Cirigliano, G.~Ecker and H.~Neufeld,
                      	Phys. Lett. {\bf B513}, 361 (2001).

\bibitem{ecker2}      	V.~Cirigliano, G.~Ecker and H.~Neufeld, 
	              	JHEP {\bf 0208}, 2 (2002).

\bibitem{bnl_2004}    	Muon g-2 Collaboration (G.W.~Bennett \ea),
		      	Phys. Rev. Lett. {\bf 92}, 161802 (2004).

\bibitem{kloe_pipi}   	KLOE Collaboration (A. Aloisio \ea),
                      	hep-ex/0407048 (2004).

\bibitem{isr}	      	S.~Binner, J.H.~Kuehn and K.~Melnikov,
                      	Phys. Lett. {\bf B459}, 279 (1999);
			G.~Rodrigo, H.~Czyz, J.H.~Kuhn and M.~Szopa,
			Eur. Phys. J. {\bf C24}, 71 (2002); see 
                      	also the later publications on the subject from
                      	these authors and collaborators.

\bibitem{babar_3pi}   	\babar\  Collaboration (B. Aubert \ea),
		      	BABAR-PUB-04/034, hep-ex/0408078 (2004).

\bibitem{dm2_3pi}	DM2 Collaboration (A. Antonelli \ea),
			Z. Phys. {\bf C56}, 15 (1992).

\bibitem{hughes}      	V.W.~Hughes and T.~Kinoshita,
                      	Rev. Mod. Phys. {\bf 71}, 133 (1999);
			T.~Kinoshita and M.~Nio,
			hep-ph/0402206 (2004).

\bibitem{cm}          	A.~Czarnecki and W.J.~Marciano,
                      	Nucl. Phys. (Proc. Suppl.) {\bf B76}, 245 (1999);
			M.~Davier and W.J.~Marciano, 
			Annu. Rev. Nucl. Part. Sci. {\bf 54}, 115 (2004).

\bibitem{krause2}     	B.~Krause, Phys. Lett. {\bf B390}, 392 (1997).

\bibitem{amuweak}     	A.~Czarnecki, W.J.~Marciano and A.~Vainshtein,
		      	Phys. Rev. {\bf D67}, 073006 (2003);
                      	A.~Czarnecki, B.~Krause and W.J.~Marciano,
                      	Phys. Rev. Lett. {\bf 76}, 3267 (1995);
                      	Phys. Rev. {\bf D52}, 2619 (1995);
		      	R.~Jackiw and S.~Weinberg, 
                      	Phys. Rev. {\bf D5}, 2473 (1972);
                      	S.~Peris, M.~Perrottet and E.~de Rafael,
                      	Phys. Lett. {\bf B355}, 523 (1995);
                      	M.~Knecht \ea, 
		      	JHEP {\bf 0211}, 003 (2002).

\bibitem{lbl_mv}	K.~Melnikov and A.~Vainshtein, 
			UMN-TH-2224-03, hep-ph/0312226 (2003).

\bibitem{rafael}      	M.~Gourdin and E.~de~Rafael, 
                      	Nucl. Phys. {\bf B10}, 667 (1969);
                      	S.J.~Brodsky and E.~de Rafael, 
                      	Phys. Rev. {\bf 168}, 1620 (1968).

\bibitem{cleo_2pi}    	CLEO Collaboration (S.~Anderson \ea),
                      	Phys.Rev. {\bf D61}, 112002 (2000).

\bibitem{opal_2pi}    	OPAL Collaboration (K. Ackerstaff \ea),
                      	Eur. Phys. J. {\bf C7}, 571 (1999).

\bibitem{bijnens}     	J.~Bijnens and P.~Gosdzinsky, 
                      	Phys. Lett. {\bf B388}, 203 (1996).

\bibitem{gj}          	S.~Ghozzi and F.J.~Jegerlehner, 
			Phys. Lett. {\bf B583}, 222 (2004).

\bibitem{md_tausf}    	M.~Davier, hep-ex/0312065 (2003).

\bibitem{pich-portoles} F.~Guerrero and A.~Pich, 
			Phys. Lett. {\bf B412}, 382 (1997);
                     	D.~G\'omez Gumm, A.~Pich and J.~Portol\'es,
                     	Phys. Rev. {\bf D62} (2000) 054014.

\bibitem{graziano}      G.~Venanzoni, private communication (Aug. 2004).

\end{thebibliography}
\end{document}